\documentclass[11pt]{article}
\usepackage{moriond,epsfig}

\def\Journal#1#2#3#4{{#1} {\bf #2}, #3 (#4)}

\def\NPB{{\em Nucl. Phys.} B}
\def\PLB{{\em Phys. Lett.}  B}
\def\PRL{\em Phys. Rev. Lett.}
\def\PRD{{\em Phys. Rev.} D}
\def\ZPC{{\em Z. Phys.} C}

\def\be{\begin{equation}}
\def\ee{\end{equation}}
\def\bea{\begin{eqnarray}}
\def\eea{\end{eqnarray}}

\def\lsim{\raise0.3ex\hbox{$\;<$\kern-0.75em\raise-1.1ex\hbox{$\sim\;$}}}
\def\gsim{\raise0.3ex\hbox{$\;>$\kern-0.75em\raise-1.1ex\hbox{$\sim\;$}}}

\begin{document}
\vspace*{4cm}
\title{SUPERSYMMETRY WITHOUT UNIVERSALITY: CP VIOLATION AND MIXING IN B MESONS}

\author{M. PIAI }

\address{SISSA-ISAS, Via Beirut 2-4, Trieste, Italy and \\
INFN,  Sezione di Trieste, Trieste, Italy}

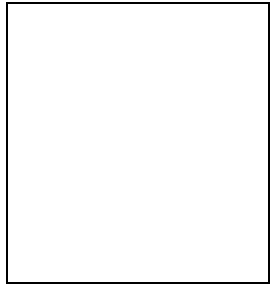
\maketitle\abstracts{
We study flavour physics and CP violation 
in the context of low energy
Supersymmetry with non-universal soft mass terms for sfermions.  
Large deviations from Standard Model predictions
are allowed in B-physics, as shown in two explicit examples. 
In particular, the consequences of these models
for $a_{J/\psi K}$ and $\Delta m_{B_s}$  are worked out.}

\section{Introduction}

The first measurements of the time-dependent 
CP asymmetries in the decays $B \rightarrow J/\psi K$ ($a_{J/\psi K}$)
marked the starting point of a long experimental program to
be carried on in next years
with the aim of 
pushing our knowledge of flavour physics in the hadronic sector
to the highest precision, in particular for CP violating observables
in B-meson physics. 
In the Standard Model (SM) all flavour
and CP violation physics is controlled by the Cabibbo-Kobayashi-Maskawa
unitary matrix (CKM), that can be parametrized in terms of three mixing
angles and just one complex phase.
Unitarity relations of the CKM can be translated in 
a graphical representation 
in the ($\bar{\rho},\bar{\eta}$) plane~\cite{Wolf}
(Unitarity Triangle, UT). At present the UT is reconstructed from 
the measured values of $\varepsilon_K$ (CP-odd), $|V_{ub}/V_{cb}|$
and $\Delta m_{B_d}$ (CP-even), and from the lower limit on $\Delta m_{B_s}$.
This allows to give definite predictions for observables
related to the angles of the UT, and
it is in principle possible 
to use this tool in the (indirect) search for new physics affecting
flavour physics, thanks to the fact that $\Delta F=2$ observables 
can receive large
contributions from virtual exchange of new heavy particles in loop 
dominated processes. Especially those observables that are theoretically
clean (i.e. free from large hadronic uncertainties), as $a_{J/\psi K}$
is, could deviate in sizable ways from the SM predictions, signaling 
an inconsistency in the pure SM Unitarity Triangle analysis.
This would be a first
evidence for new physics before LHC starts operating,
 in particular for Supersymmetry (SUSY), which
seems to be the most promising extension of Standard Model with
 interesting low energy
features. 
In spite of this, it has been understood that without 
a completely new flavour structure in the soft terms of the SUSY 
Lagrangian it is vey unlikely to have such a clear
 signature: for large classes
of supersymmetric models with no new sources of flavour violation
(C-MSSM for example) the deviations of CP-odd observables from SM prediction
 are so small~\cite{Demir:2000ky}
that the experimental precision reachable makes very difficult to detect them.

We construct two explicit examples~\cite{Masiero:2000ni}
 of supersymmetric extensions of
SM with non-universal soft terms, in which large deviations are predicted 
from the standard UT, while  a dynamical motivation is provided
for the peculiar flavour structure needed in the sfermionic sector to 
avoid exceedingly large contributions to Flavour Changing Neutral Current 
(FCNC), that could arise in a general SUSY model.
In both these models we are allowed to work with Mass Insertion Approximation
(MIA) and we focus our attention on soft sfermion masses neglecting
 the effects of the LR insertions, that we assume to be very small, 
in order not to exceed bounds coming from EDMs, $\varepsilon^{\prime}/\varepsilon$ and $b\rightarrow s \gamma$ transitions.

\section{Non-universal singlet masses}

As a first example, we consider a model inspired to Type-I string 
theory~\cite{typeI}. Squark doublets have universal masses,
while singlets have not:
\be
\left\{\begin{array}{ccc}
(\tilde{m}^2_Q)_{ii}& =& m^2_{3/2} (1-\frac{3}{2}\cos^2\theta(1-\Theta^2_1))\\
(\tilde{m}^2_{d})_{ii}& =& m^2_{3/2} (1-3\cos^2\theta\Theta^2_i)\end{array}\right. .
\ee
$\theta$ and $\Theta_i$ are goldstino angles, that we can treat as free parameters, and $m_{3/2}$ is the gravitino mass.
In the basis where squarks are diagonal $V_{CKM} = (V^{U})^{\dagger} V^D$,
being $V^{U,D}$ the unitary matrices rotating left fields in the 
diagonalization of the Yukawa couplings.
Let us assume that $V_{CKM}$ is real (extreme situation, not realistic)
 and that the $V^{U,D}$ matrices have the same structure as the CKM one 
(conservative assumption). The re-phasing needed to make the CKM
real leaves a 
set of new observable phases in the $V^{U,D}$ matrices (denoted here
$\varphi_i$, $i=1, \cdots, 3$), which will appear
 in the expression of the mass insertions, written in the basis in which
fermions are diagonal:
\bea
V^{D}\simeq\ \left(\begin{array}{ccc} 1 - \lambda^2/2 & \lambda \ e^{i \varphi_1}  
& A \lambda^3 e^{i\varphi_2} \\ 
-\lambda \ e^{-i\varphi_1} & 1 - \lambda^2/2 & A \lambda^2 \ e^{i\varphi_3}\\ 
A \lambda^3 \ ( e^{- i(\varphi_1+\varphi_3)}-\rho e^{-i\varphi_2})  & -A \lambda^2 \ e^{- i\varphi_3}  & 1 
\end{array}\right),\\
(\delta^{d}_{RR})_{i j}= \frac{1}{\tilde{m}^{2}_{\tilde{q}}} \Big(
(\tilde{m}_{d_{2}}^2 - \tilde{m}_{d_{1}}^2 )\ V^{D}_{i 2} {V^{D}_{j
2}}^*\  +\ (\tilde{m}_{d_{3}}^2 - \tilde{m}_{d_{1}}^2 )\ V^{D}_{i 3} 
{V^{D}_{j 3}}^* \Big).
\eea
$\lambda$, $\rho$ and $A$ are the parameters of the Wolfenstein 
parameterization of CKM~\cite{Wolf}.
After the SUSY running down to the electroweak scale (in the hypothesis of 
gluino dominance), we find:
\bea
\Im(\delta^{d}_{RR})_{1 2}&\simeq& \frac{\cos^2 \theta (\Theta_1^2 - \Theta_2^2)}{ 7 \sin^2 \theta} \ 
\lambda \ \sin{\varphi_1} \\
&\simeq&  0.03 (\Theta_1^2 -\Theta_2^2) \sin \varphi_1
\lsim 0.0032,
\eea 
for $\sin\theta \sim 0.7$. So it is possible to fully saturate $\varepsilon_K$
with SUSY. On the contrary the 13 insertion turns out to be proportional 
to $A\lambda^3/6 \sim 10^{-3}$, while the saturation limit from B physics is
around $0.098$~\cite{Gabbiani:1996hi}. This means that SUSY contributes 
negligibly to $B_d$ mixing, the UT collapses to a line due to the vanishing 
of the phase in the CKM, and $a_{J/\psi K}$ is predicted to vanish.
Actually this is not compatible with CP-even observable measurements, 
that predict a non-zero phase in the CKM, which could however be
much smaller than the one predicted by Standard UT (in which $\varepsilon_K$
plays a crucial role). Values as low as $a_{J/\psi K} \sim 0.2 \div 0.3$
could be accommodated, still satisfying all other possible experimental 
constraints.

\section{Non-abelian flavour symmetries}

To have a sizable contribution to $B$-meson oscillation parameters,
without exceeding the bounds coming from the $K^0-\bar{K}^0$ system,
one can impose a flavour symmetry to forbid large 12 insertions. This
is what happens in a class of supersymmetric models with non-abelian horizontal
flavour symmetry, such as $U(2)$ or $SU(3)$. These models
gave rise to a lot of interest in recent years due to the constraints they
impose on fermionic mass matrices, which allow to fit the UT, giving 
predictions compatible with experiments for a large variety of observables,
both in the hadronic and in the leptonic sector of the SM, in particular 
for realizations embedded in Grand Unified Theories (GUT)~\cite{flav}.
In the context of Supergravity mediated SUSY breaking, these symmetries
impose analogous constraints also on the sfermionic textures, that can 
give important contributions to both kaon and B-meson physics 
(for $\Delta F=2$ observables),
changing drastically the fit of the UT based only on fermionic textures.
This would lead to large deviations from SM for the $B_s$ mass difference,
and for the time dependent CP asymmetries in $B_d \rightarrow J/\psi K$.
As an instance of this, let us consider a model based on $SU(3)$, very similar 
to others discussed in the literature~\cite{Berezhiani:2001cg}. 
In such a model
$SU(3)$ is broken by the vacuum expectation values (VEVs) of a set of 
SM singlets (flavons) carrying $SU(3)$ quantum numbers,
and matter fields
are assigned to transform as a triplet. Symmetry breaking is communicated to
matter through heavy degrees of freedom, resulting in suppression 
factors for the couplings proportional to the ratio between the VEVs and this 
heavy mass scale.
In our specific case, we find that the textures for hadronic matter
fields, neglecting small higher order terms,  are the following (at GUT scale):
\be
\begin{array}{cc}
\label{down}
M_d=m^D  \left(\begin{array}{ccc} 0 & \epsilon^{\prime} & 0 \\
       - \epsilon^{\prime} & c \eta                 & b \epsilon \\    
       0                   & \epsilon          & \eta  
\end{array}\right) \  \ & \  \
\tilde{m}^2_Q = m^2_{3/2}
\left(
\begin{array}{ccc}
1& 0 & \alpha\epsilon\epsilon^{\prime} \\
0 & 1+\lambda \epsilon^2 & \beta \epsilon \eta \\
\alpha^*\epsilon\epsilon^{\prime} & \beta^*\epsilon\eta & r_3
\end{array} \right) \\
 \  \ &  \  \   
\tilde{m}^2_d = m^2_{3/2} 
\left(
\begin{array}{ccc}
1& 0 & \alpha^{\prime}\epsilon\epsilon^{\prime} \\
0 & 1+ \lambda^{\prime} \epsilon^2  & \beta^{\prime} \epsilon \eta \\
{\alpha^{\prime}}^*\epsilon\epsilon^{\prime} & {\beta^{\prime}}^{ \ast} \epsilon \eta & r_3
\end{array} \right) 
\end{array} 
\ee
Here $c \simeq m_c/m_t$, $r_3 \equiv \tilde{m}^2_3/\tilde{m}^2_1$ is 
the ratio between the masses of third and first family ($SU(3)$ breaking 
is large), $m^D$ is proportional to the mass of the
bottom quark, while $1>\eta>\epsilon>\epsilon^{\prime}$
are suppression factors due to symmetry reasons. The other 
coefficients are $\cal{O}$$(1)$ couplings, apart from $b$ which can be 
small due to the presence of the adjoint representation flavon. Up-type
mass matrices are assumed to be diagonal.

The structure of the fermionic textures is such that the Jarskog 
determinant~\cite{Jarlskog} is small: a first prediction of such 
a model is that the phase in the CKM can be neglected,
assuming all parameters entering fermionic matrices to be 
real. In this way the UT degenerates to a line and
the best fit of
the fermionic textures to
the masses of quarks, and to the elements of the CKM determined by
tree level processes ($|V_{ub}|$, $|V_{cb}|$ and $|V_{us}|$), is 
obtained with $\bar{\rho}<0$. This leads to a 
SM contribution to $\Delta m_{B_d} \sim 1$ ps$^{-1}$, a factor of two larger 
than the experimental value (SUSY and NLO-QDC running are included), 
so requiring a large SUSY contribution.

Then we compute the mass insertions and the 
$\Delta F=2$ amplitudes. We perform a scanning of the parameter 
space available, imposing the experimental values of
$\varepsilon_K$
and $\Delta m_{B_d}$,
and extract the prediction for $\Delta m_{B_s}$ and $a_{J/\psi K}$.
In doing this we notice that not all the free parameters
are really relevant for the analysis itself:
$\alpha$ and 
$\alpha^{\prime}$ could be safely neglected, while
  loop factors make
only a combination of $\beta$ and
$\beta^{\prime}$ to be important.
 In particular this causes a certain degree
of correlation between the SUSY contributions to 
$B_d$ and $B_s$ mixing.
All experimental constraints are satisfied, and we find that 
large discepancies  from the SM~\cite{Ciuchini}
 are possible in large samples of the 
parameter space. In particular the model allows for every possible value 
of the time-dependent CP asymmetries, also in $B_s\rightarrow J/\psi \phi$.

Analogous results can be obtained in other models, based for instance
on a $U(2)$ symmetry, although the reality of the CKM is quite a 
peculiar feature, that needs strong constraits on the couplings of the model.

\begin{figure}
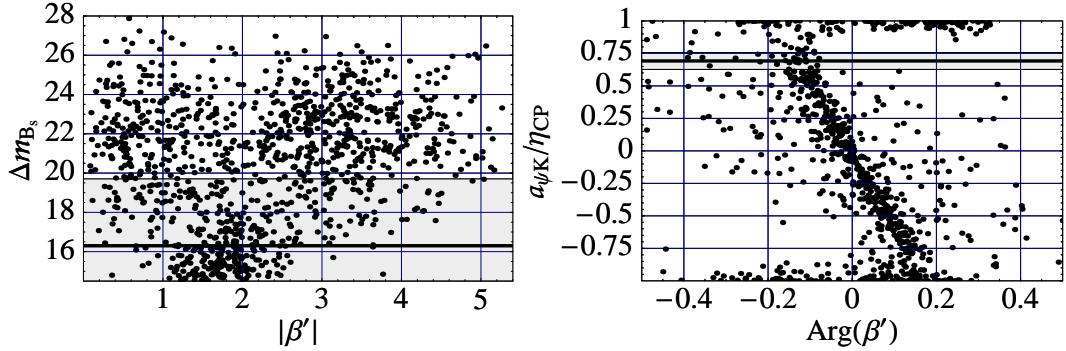

\begin{center}
\epsfig{file=dmbsbbldm.epsi,width=0.42\linewidth}
\epsfig{file=apsikbba.epsi,width=0.45\linewidth}
\caption{Scatter plots in $SU(3)$ model, compared to SM 
predictions (in grey)
.}
\end{center}
\end{figure}

\section{Conclusions}

Non-universal realizations of SUSY can give large contributions to 
$\Delta F=2$ observables, making them distinguishable from SM and 
making possible an indirect discovery of SUSY in precision measurements
in B-meson physics.
In particular CP violation could receive large, or even dominant
contributions from non-universal SUSY.
A realistic model, in which B-physics is strongly affected by SUSY,
needs a non-trivial flavour structure,
that can be provided by non-abelian flavour symmetries, where the 
effect of the sfermionic sector must be taken into account in
the UT fit. 

\section*{Acknowledgements}

I would like to thank the organizers for the pleasant atmosphere
of the Conference and for financial support.  
Special thanks to A. Masiero, A. Romanino, L. Silvestrini and 
O. Vives, for the fruitful collaboration this talk is based on.


\section*{References}
\begin{thebibliography}{99}


\bibitem{Wolf} 
L.~Wolfenstein,
\Journal{\PRL}{51}{1945}{1983};
A.~J.~Buras {\it et al.},
\Journal{\PRD}{50}{3433}{1994}.

\bibitem{Demir:2000ky}
D.~A.~Demir {\it et al.},
\Journal{\PRD}{61}{075009}{2000};
A.~Ali {\it et al.},
{\tt hep-ph/0002167};
A.~J.~Buras {\it et al.},
\Journal{\PLB}{501}{223}{2001}.

\bibitem{Masiero:2000ni}
A.~Masiero {\it et al.},
{\tt hep-ph/0012096};
{\tt hep-ph/0104101}.

\bibitem{typeI}
L.E.~Iba\~nez {\it et al.},
\Journal{\NPB}{553}{43}{1999};
M.~Brhlik {\it et al.},
\Journal{\PRL}{84}{3041}{2000}.




\bibitem{Gabbiani:1996hi}
F.~Gabbiani {\it et al.},
\Journal{\NPB}{477}{321}{1996}.

\bibitem{flav}
see for instance: 
A.~Pomarol {\it el al.}
\Journal{\NPB}{466}{3}{1996};
R.~Barbieri {\it et al.},
\Journal{\PLB}{377}{76}{1996};
R.~Barbieri {\it et al.},
\Journal{\NPB}{559}{17}{1999}.

\bibitem{Berezhiani:2001cg}
Z.~Berezhiani and A.~Rossi,
\Journal{\NPB}{594}{113}{2001}.

\bibitem{Jarlskog}
C.~Jarlskog, 
\Journal{\PRL}{55}{1039}{1985};
\Journal{\ZPC}{29}{491}{1985}.

\bibitem{Ciuchini}
M.~Ciuchini {\it et al.},
{\tt hep-ph/0012308}.


\end{thebibliography}
\end{document}